\documentclass[
twocolumn, 
 amsmath,amssymb,
 aps,
prl,
superscriptaddress
]{revtex4-2}

\usepackage{graphicx}
\usepackage{dcolumn}
\usepackage{bm}
\usepackage[colorlinks=true, allcolors=blue]{hyperref}
\usepackage{braket}
\usepackage{xcolor}
\usepackage{bbding}
\usepackage{natbib}
\usepackage{gensymb}
\usepackage{balance}
\usepackage{comment}
\usepackage{float}  
\usepackage{tikz}
\usepackage{pgfplots}
\pgfplotsset{compat=1.18}
\usepackage{lipsum}  
\usepackage{tikz}
\usetikzlibrary{matrix, positioning, shapes.geometric, arrows.meta, decorations.pathreplacing, calligraphy}

\begin{document}
\preprint{APS/123-QED}

\title{Thermalization in a Height-Conserving Quantum Dimer Model}

\author{Junsheng Feng}
\affiliation{%
School of Physics and Material Engineering, Hefei Normal University, \\
Hefei 230601, P. R. China 
}%
\author{Jie Ren}
\email{J.Ren@leeds.ac.uk}
\affiliation{School of Physics and Astronomy, University of Leeds, Leeds LS2 9JT, United Kingdom}

\author{Zheng Yan}
\email{zhengyan@westlake.edu.cn}
\affiliation{%
Department of Physics, School of Science and Research Center for
Industries of the Future, Westlake University, Hangzhou 310030, China 
}%
\affiliation{
Institute of Natural Sciences, Westlake Institute for Advanced Study, Hangzhou 310024, China
}
\date{\today}

\begin{abstract}
Strongly constrained quantum systems, in which local rules forbid most configurations, play a central role in condensed matter and lattice gauge theory. Their thermalization is often thought to be delicate: extensive conservation laws and dynamically frozen states can shatter the Hilbert space into many disconnected sectors. A natural question is whether, once the frozen states are removed, the dynamics within a single sector still thermalizes.
We address this in the height-conserving quantum dimer model on the square lattice, whose local plaquette flips conserve an emergent height field. Resolving the winding numbers, the four sublattice heights, and lattice momentum, we isolate the dominant connected Krylov component of each fragmented sector and analyze its spectral statistics, entanglement, and connectivity.
The two standard chaos diagnostics then show different behavior: across momentum sectors the level-spacing statistics range from near-Poisson to Wigner-Dyson, yet in every sector the eigenstate entanglement entropy collapses onto a narrow, dome-shaped curve characteristic of eigenstate thermalization. Only a handful of low-entanglement outliers interrupt this thermal pattern, in selected sectors. Thus, strong kinematic constraints can lead to a situation where spectral correlations and eigenstate thermalization need not follow the same universal signatures—a manifestation of constrained quantum chaos.

\end{abstract}

\maketitle

\section{Introduction}
Quantum chaos in many-body systems is traditionally diagnosed by two complementary probes: spectral statistics and eigenstate thermalization. In generic nonintegrable systems without fine-tuned conservation laws, the level spacing distribution follows Wigner–Dyson statistics, indicating strong level repulsion, and the eigenstates satisfy the eigenstate thermalization hypothesis (ETH), meaning that individual eigenstates reproduce thermal expectation values of local observables \cite{Bohigas1984,Deutsch1991,Srednicki1994,rigol2008thermalization,DALESSIO2016,KimIkedaHuse2014,Gogolin2016}. Conversely, integrable systems and many-body localized phases exhibit Poisson level statistics, reflecting uncorrelated energy levels, and their eigenstates typically violate ETH, showing area-law entanglement or large fluctuations \cite{BerryTabor1977,Basko2006,PalHuse2010,NandkishoreHuse2015,AbaninRMP2019,BauerNayak2013,Moudgalya2018,Khemani2020,Sala2020}. This dichotomy has become a cornerstone of our understanding of thermalization in isolated quantum systems.

However, a growing number of constrained lattice models have revealed more nuanced scenarios. Strong local constraints, such as dipole conservation, gauge symmetries, or Hilbert-space fragmentation, can produce spectral statistics that are neither purely Poisson nor fully Wigner–Dyson \cite{Moudgalya2021,theveniaut2020transition,pietracaprina2021probing,mace2019many}, while simultaneously hosting eigenstates that appear thermal or, in other cases, exhibit many-body scarring \cite{Bernien2017,TurnerNatPhys2018,Turner2018,Serbyn2021,Lin2020,Moudgalya2018a,Moudgalya2022}.
In such systems, the conventional link between level repulsion and ETH may be broken. Yet, a key question remains: after fully resolving all conserved quantities and removing the inert frozen states, does the residual dynamics within a connected sector still lead to thermalization? Even in globally fragmented systems, a given Krylov subspace might still support ETH-like eigenstates if the effective Hamiltonian within it is sufficiently mixing. Understanding this interplay is essential for a complete theory of quantum thermalization under constraints.

In this work, we address this problem by studying a typical highly constrained model -- the height-conserving quantum dimer model (hQDM) on the square lattice under periodic boundary conditions \cite{Yan22}. In addition to the commonly two winding numbers of the quantum dimer model~\cite{Moessner2010,yan2022global}, the conserved height fields contribute four additional conserved quantities --- the respective total heights on the four sublattices are invariant under the quantum dynamics of this model. In other words, the model is defined by local plaquette-flip dynamics that conserves an emergent height field and its associated winding numbers, resulting in a set of six conserved quantities. The Hilbert space fragments into sectors labelled by these quantities; however, within each sector, after removing frozen states, there remains a dominant connected Krylov component that contains the overwhelming majority of states. The excluded components are dynamically frozen and do not affect thermalization. We further resolve lattice translation symmetry to obtain momentum blocks and restrict our analysis to these dominant Krylov components, allowing us to probe the intrinsic spectral and eigenstate properties of the constrained dynamics without trivial degeneracies or disconnected frozen parts.

To explore the thermalization properties in highly constrained systems, we analyze three representative momentum sectors of the $8\times 8$ lattice. We observe that the level statistics vary between sectors, with some showing Wigner–Dyson-like behavior and others closer to Poisson statistics. Remarkably, in all sectors, the many-body eigenstates exhibit a smooth, dome-shaped energy-resolved entanglement entropy, indicating that the majority of states remain thermal despite the presence of a few low-entanglement outliers. While the limited system size prevents a systematic finite-size scaling, these findings support the view that, once frozen states are removed, the Hamiltonian subspaces remain largely connected and thermalizable, consistent with the eigenstate thermalization hypothesis in strongly constrained systems.

\section{Model}
As a typical highly-constrained model, we study the height-conserving quantum dimer model on the square lattice, where strong local constraints coexist with fully local dynamics~\cite{Yan22}. Compared with the conventional square-lattice quantum dimer model, the defining feature of the hQDM is the conservation of total height-field in each subspace under all allowed resonance processes. This additional structure severely restricts dynamical pathways in Hilbert space, making the model ideally suited for investigating quantum chaos and thermalization in constrained subspaces.

We consider the Hamiltonian
\begin{equation}
\hat{H} = \hat{H}_{\mathrm{res}} + \hat{H}_{\mathrm{diag}},
\end{equation}
with resonating terms
\begin{equation}
\hat{H}_{\mathrm{res}} = -t \sum_{plaq} \left(
|\vcenter{\hbox{\includegraphics[width=1cm]{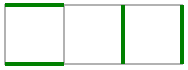}}}\rangle
\langle\vcenter{\hbox{\includegraphics[width=1cm]{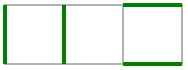}}}| + \text{H.c.}
\right),
\end{equation}
and diagonal terms
\begin{equation}
\hat{H}_{\mathrm{diag}} = v \sum_{plaq} \left(
|\vcenter{\hbox{\includegraphics[width=1cm]{hv_plaq.pdf}}}\rangle
\langle\vcenter{\hbox{\includegraphics[width=1cm]{hv_plaq.pdf}}}|
+
|\vcenter{\hbox{\includegraphics[width=1cm]{vh_plaq.pdf}}}\rangle
\langle\vcenter{\hbox{\includegraphics[width=1cm]{vh_plaq.pdf}}}|
\right).
\end{equation}
The sums are run over all vertically or horizontally aligned next-to-nearest-neighbor plaquette pairs. The resonance term flips a pair of parallel plaquettes, while the diagonal term assigns an energy cost to flippable configurations.

As in conventional square-lattice quantum dimer model, the fully-packed-dimer constraint introduces a compact U(1) gauge field which leads to a conservation of winding numbers under local fluctuations~\cite{Moessner2010}. Similarly, the hQDM inherits two topological winding numbers,
\begin{equation}
W_x = \frac{h(L,y)-h(0,y)}{L}, \qquad
W_y = \frac{h(x,L)-h(x,0)}{L},
\end{equation}
which are conserved under all local dynamics. Here the $W_x$ and $W_y$ are the winding numbers along the $x$ and $y$ directions, $h(x,y)$ is the height field defined on the plaquette $(x,y)$ \cite{Yan22} and $L$ is the lattice length.

\begin{figure}[!htbp]
        \centering
        \includegraphics[width=0.5\textwidth]{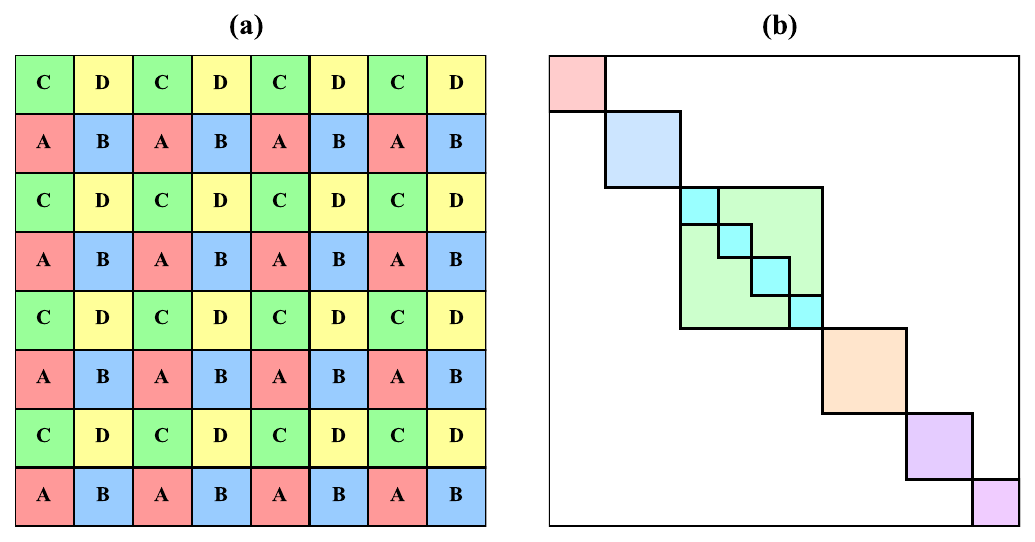}
        \caption{
(a) The four dual sublattices \textit{A}, \textit{B}, \textit{C}, \textit{D} used in the construction of the height variables.
(b) Schematic illustration of hierarchical block-diagonal structure of a fragmented Hamiltonian.
}
        \label{fig:height_field_plot}
    \end{figure}

In addition, the restricted geometry of the pairwise plaquette flips enforces four further conserved quantities: the total heights on each of the four sublattices,
\begin{equation}
I_X = \sum_{\mathbf{r}\in X} h(\mathbf{r}), \qquad X=A,B,C,D,
\end{equation}
as illustrated in Fig.~\ref{fig:height_field_plot} (a). Together, the six conserved quantities ${W_x,W_y,I_A,I_B,I_C,I_D}$ impose strong kinematic constraints on the dynamics.
Being finite in number, these conservation laws block-diagonalize $\hat{H}$ but do not render the model integrable. The constraints suppress spectral mixing, and some frozen states further fragment the sectors~\cite{Yan22}.

Upon resolving the conserved quantities and lattice translation symmetry, the Hamiltonian acquires a block-diagonal structure as shown in the Fig.\ref{fig:height_field_plot} (b),
$\mathcal{H} = \bigoplus_{\beta} \mathcal{H}^{(\beta)}$,
where $\beta \equiv (W_x,W_y,I_A,I_B,I_C,I_D,k_x,k_y)$, $k_x$ and $k_y$ are the momentum. Each block $\mathcal{H}^{(\beta)}$ corresponds to a fixed symmetry sector. As mentioned above, there may be frozen states further fragmenting the subspaces, thus we need to do the simulation in Krylov spaces, forming the basis for our subsequent analysis of spectral statistics and eigenstate properties.

\section{Hilbert-space construction and symmetry resolution}
The fully packed constraint together with the conserved height-field quantities induces a highly structured Hilbert space. To faithfully analyze spectral and eigenstate properties, it is essential to explicitly resolve these constraints and construct symmetry-adapted Hamiltonian blocks.

We begin from the complete set of fully packed dimer coverings on an $L\times L$ square lattice with periodic boundary conditions. From this space, we first eliminate frozen configurations, i.e., states that do not contain any flippable plaquette pairs of hQDM and are therefore exact eigenstates of $\hat{H}$ with zero kinetic energy. Since such states are dynamically inert and trivially localized, they are excluded from all subsequent analyses.

Each remaining configuration is uniquely labeled by the six conserved quantities
$\alpha = (W_x, W_y, I_A, I_B, I_C, I_D)$,
which define disjoint Hilbert-space sectors. Note that $\alpha$ does not contain momentum $k_x$ and $k_y$ which distinguishes it from $\beta$. The full Hilbert space thus decomposes as
$\mathcal{H} = \bigoplus_{\alpha} \mathcal{H}^{(\alpha)}.$
Within each sector, basis states are connected by sequences of allowed pair-plaquette flips generated by $\hat{H}_{\mathrm{res}}$.

For each symmetry sector $\mathcal{H}_{\alpha}$, we explicitly construct the Hamiltonian matrix in the real-space dimer-covering basis. We then analyze the connectivity graph induced by the nonzero off-diagonal matrix elements and identify its connected components. In all cases, a single giant connected component exhausts the overwhelming majority of basis states, which we take as the relevant Krylov subspace for subsequent analysis. This dominant connected component is the Krylov subspace we analyze. As a consequence, the largest connected component does not contain the domain-wall states reported in Ref.~\cite{Yan22}, since those states belong to small dynamically disconnected Krylov subspaces. Our analysis therefore focuses exclusively on the ergodically connected part of each symmetry sector. See Appendix G for a detailed description of the graph construction and connectivity analysis.

Next, we examine whether lattice translation symmetry is preserved within a given $(W_x,W_y,I_A,I_B,I_C,I_D)$ sector. When translations by lattice vectors commute with the Hamiltonian within the sector, we further resolve momentum quantum numbers by projecting onto Bloch sectors labeled by crystal momentum $\mathbf{k}=(k_x,k_y)$.

Concretely, for each real-space basis state $|c\rangle$, we generate its translational orbit
$|\mathbf{k}, c\rangle = \frac{1}{\sqrt{N_c}} \sum_{\mathbf{r}}
e^{i\mathbf{k}\cdot\mathbf{r}}  T_{\mathbf{r}} |c\rangle$,
where $T_{\mathbf{r}}$ denotes lattice translation and $N_c$ is a normalization factor accounting for stabilizers. This procedure yields momentum-resolved Hamiltonian blocks
\begin{equation}
\hat{H} = \bigoplus_{\beta} \hat{H}^{(\beta)}, \qquad
\beta = (W_x,W_y,I_A,I_B,I_C,I_D,k_x,k_y)
\end{equation}

All spectral statistics and eigenstate diagnostics presented in the following are performed within these fully symmetry-resolved momentum sectors.

\section{Diagnostics}
To characterize quantum chaos, thermalization, and possible nonergodic structures within each symmetry-resolved sector, we employ complementary diagnostics based on spectral statistics, eigenstate properties, and Krylov-space graph structure.

\textbf{\textit{Spectral statistics.—}}
For each momentum-resolved block, we compute the distribution of adjacent energy-gap ratios
\begin{equation}
r_n = \frac{\min(\delta_n,\delta_{n+1})}{\max(\delta_n,\delta_{n+1})},\quad
\delta_n = E_{n+1}-E_n ,
\end{equation}
and its average $\langle r\rangle$ over the bulk of the spectrum \cite{Oganesyan2007,Atas2013}. 
For reference, Poisson statistics yield $\langle r\rangle \approx 0.386$, while Wigner--Dyson statistics of the Gaussian orthogonal ensemble (GOE) give $\langle r\rangle \approx 0.536$. For the Gaussian unitary ensemble (GUE), $\langle r\rangle \approx 0.603$; because the Hamiltonian is real in the dimer basis, a momentum block follows GOE when a combined reflection and time-reversal symmetry survives within the sector and GUE otherwise. 
In addition, we compute the spectral form factor (SFF),
\begin{equation}
K(t)=\left\langle \left| \sum_n e^{-iE_n t} \right|^2 \right\rangle ,
\end{equation}
after local spectral unfolding, and extract the slope of the linear ramp regime as a further probe of long-range spectral correlations \cite{Cotler2017,Chan2018}. In practice, we compute the time-averaged SFF without performing any ensemble averaging, as detailed in Appendix E.

\textbf{\textit{Eigenstate entanglement.—}}
To diagnose eigenstate thermalization, we compute the bipartite von Neumann entanglement entropy
\begin{equation}
S = -\mathrm{Tr}_A \rho_A \ln \rho_A,
\end{equation}
for each many-body eigenstate under an equal spatial bipartition.
ETH is associated with a smooth, narrow distribution of $S$ as a function of energy density, forming a dome-shaped structure, whereas isolated low-entanglement outliers signal atypical nonthermal eigenstates \cite{Deutsch1991,Srednicki1994,rigol2008thermalization,Moudgalya2018a}.

\textbf{\textit{Krylov-space graph diagnostics.—}}
We represent each Hamiltonian block as a graph whose nodes correspond to basis states and whose edges correspond to nonzero off-diagonal matrix elements.
From this graph we compute:
(i) the number of connected components,
(ii) the degree distribution and its relative standard deviation,
and (iii) the smallest eigenvalues of the graph Laplacian.
A single connected component with a finite spectral gap of the Laplacian indicates the absence of Hilbert-space fragmentation \cite{Sala2020,Khemani2020}.

To probe a coherent structure in the complex hopping amplitudes, we additionally define a phase-order parameter
\begin{equation}
\Phi = \left| \frac{1}{N_b}\sum_{(i,j)} e^{i\arg H_{ij}} \right|,
\end{equation}
where the sum runs over all nonzero off-diagonal bonds.
Large $\Phi$ signals the coherent phase alignment characteristic of structured dynamics, while $\Phi \approx 0$ indicates random-matrix-like phases.

\textbf{\textit{Combined interpretation.—}}
To characterize the dynamics of the height-conserving quantum dimer model, we employ a combination of complementary diagnostics. Level-spacing distributions and average adjacent-gap ratios ($\langle r\rangle$) identify the degree of level repulsion in each momentum subsector, while time-averaged spectral form factors quantify correlations and the presence of ramp behavior in the bulk spectrum. Bipartite von Neumann entanglement entropy provides an independent measure of eigenstate thermalization, revealing smooth, dome-shaped energy-resolved bands. Graph-theoretic analysis of the Krylov subspace connectivity and the phase coherence of off-diagonal Hamiltonian elements further distinguishes the role of local constraints from structural effects such as Hilbert-space fragmentation. Taken together, these tools form a unified framework that allows us to systematically analyze multiple momentum subsectors and sectors, and to identify the emergence of constrained quantum chaos, the coexistence of thermal and scar states, and momentum-dependent variations in spectral statistics.

\section{Results and discussion}
We now apply the above diagnostics to symmetry-resolved sectors of hQDM. For a square lattice, the maximum system size feasible for exact diagonalization (ED) is $8\times8$.  To minimize parameter-dependent effects on spectral statistics, we fix $v/t=-0.5$, which lies away from special fine-tuned points such as the Rokhsar–Kivelson point and classical limits, and corresponds to a generic interacting regime with strong resonance dynamics. Although all sectors considered are fully connected within their dominant Krylov component, we find striking sector-to-sector variations in spectral statistics and eigenstate structure.  In particular, some momentum sectors exhibit ETH-consistent entanglement with no visible low entanglement outliers, while others host isolated atypical eigenstates with anomalously low entanglement entropy, indicative of many-body scarring.

To elucidate this dichotomy, we focus on three representative classes of sectors: one with translational symmetry along the $x$-direction, one along the $y$-direction, and the last one possessing translational symmetries along both the $x$- and $y$-directions. The translation unit is two sites: single-site translations permute the sublattice charges $I_X$, so only shifts by two preserve a given sector.

\textbf{\textit{Case}-I:}
In the sector with $x$-direction translational symmetry resolved, extracting the largest connected Krylov component (total Krylov-space dimension 18592) yields four momentum subsectors 
$k_x=(0, \pi/2, \pi, 3\pi/2)$, each of dimension 4648. 

\begin{widetext}
\begin{center}
    \begin{figure}[H]
        \centering
        \includegraphics[width=0.8\textwidth]{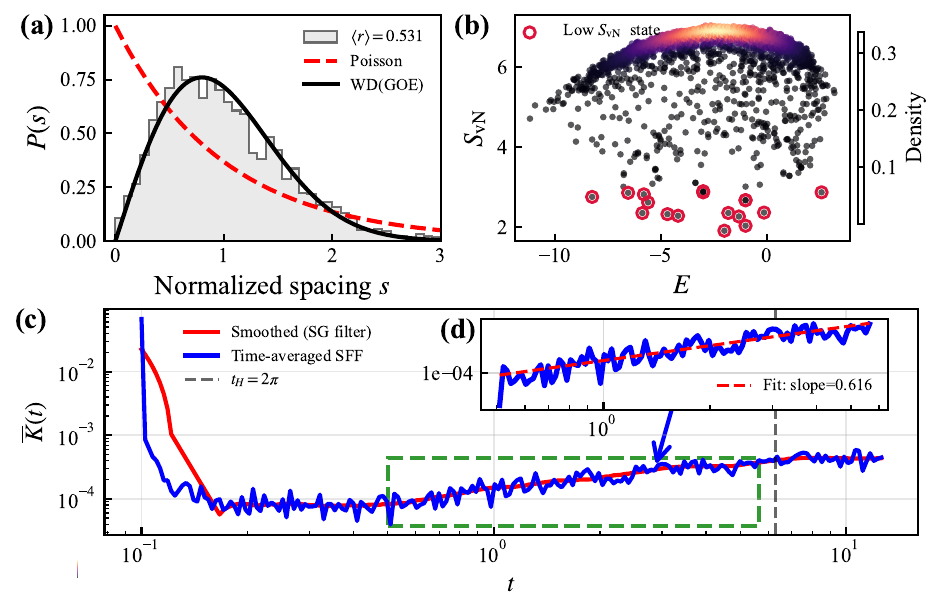}
        \caption{
Spectral and entanglement analysis for the $x$-translation symmetric sector with parameters $\alpha= (W_x = 0 , W_y = 0 , I_A = 0 ,I_B = 8  ,I_C =  60 ,I_D = 8 )$ and  momentum $\mathbf{k} = (0,\pi/2)$ . (a) Level-spacing histogram $P(s)$ with unfolded eigenvalues, showing good agreement with the Wigner-Dyson (GOE) distribution (black solid line) and comparison to the Poisson distribution (red dashed line);  (b) Bipartite von Neumann entanglement entropy $S_{vN}$ versus energy $E$, with the density estimated via a Gaussian KDE (colorbar) and low-entanglement states highlighted by red circles below the main thermal dome. (c) Time-averaged spectral form factor (SFF, blue) and smoothed SFF using Savitzky–Golay filter (red). (d) Zoomed-in view of the ramp region of the SFF (highlighted in green in panel c), showing a linear fit (red dashed line) with slope $0.616$, indicating partial level repulsion in the bulk spectrum.
}
        \label{fig:svn1}
    \end{figure}
\end{center}
\end{widetext}

FIG. \ref{fig:svn1} illustrates the spectral and entanglement properties of the $k_x= \pi/2$ subsector. Panel (a) shows the level-spacing distribution $P(s)$ of the unfolded eigenvalues, with the histogram following the Wigner–Dyson form for the bulk spectrum (black solid line) and the Poisson distribution shown for comparison (red dashed line). The average adjacent-gap ratio $\langle r\rangle = 0.531$ indicates significant level repulsion. Panel (b) presents the bipartite von Neumann entanglement entropy $S_{vN}$ as a function of energy $E$, with the density estimated via a Gaussian kernel; most eigenstates form a smooth, dome-shaped distribution, characteristic of thermal behavior, while a small number of low-entanglement scar states appear below the main thermal band (highlighted with red circles). Panel (c) displays the time-averaged spectral form factor (SFF, blue curve) and a smoothed version (red curve) obtained via a Savitzky–Golay filter, capturing the decay at short times and the onset of the ramp. Panel (d) zooms into the ramp region of the SFF (highlighted in green in panel c), with a linear fit (red dashed line) quantifying the slope of correlation growth in the bulk spectrum. Together, these panels demonstrate that the $k_x=\pi/2$ subsector exhibits spectral signatures of quantum chaos, while the majority of eigenstates remain thermal, and isolated scar states coexist with the bulk.

To better understand this behavior, we examine all four momentum subsectors from resolving translational symmetry along the $x$ direction: $k_x =(0,\pi/2,\pi,3\pi/2)$, each of equal dimension (4648) within the largest connected Krylov component (dimension 18592). The connectivity structure induced by nonzero off-diagonal Hamiltonian matrix elements is essentially identical across all four subsectors:  each forms a single connected graph with nearly identical mean degree  (denoted by the coordination number $z$, representing the average number of connections per node; here $\langle z\rangle \approx 5$) and relative degree fluctuations ($\mathrm{std}(z)/\langle z\rangle \approx 0.34$), and the graph Laplacian spectrum contains exactly one zero mode, confirming that this Krylov component is connected and structurally similar across subsectors. Despite this structural equivalence, the spectral statistics exhibit a pronounced momentum dependence: $k_x =(\pi/2,3\pi/2)$ display Wigner–Dyson-like correlations ($\langle r \rangle \approx 0.53$), whereas $k_x = 0$ and $k_x = \pi$ show weaker correlations ($\langle r \rangle \approx 0.43$), closer to Poisson. Nevertheless, all subsectors exhibit smooth, dome-shaped distribution of entanglement entropy and the SFF behavior is also indicative of thermalization, with only a small number of low-entanglement scar states appearing in the center of the spectrum. These observations demonstrate that momentum projection can strongly modulate spectral correlations without altering Hilbert-space connectivity, and that thermalization occurs across all sectors, with level statistics varying due to momentum-dependent phase coherence.

\textbf{\textit{Case}-II:} 
In the sector with $y$-direction translational symmetry resolved, extracting the largest connected Krylov component (total dimension 8760) yields four momentum subsectors $k_y=(0, \pi/2, \pi, 3\pi/2)$, each of dimension 2190.

\begin{widetext}
\begin{center} 
    \begin{figure}[H]
        \centering
        \includegraphics[width=0.8\textwidth]{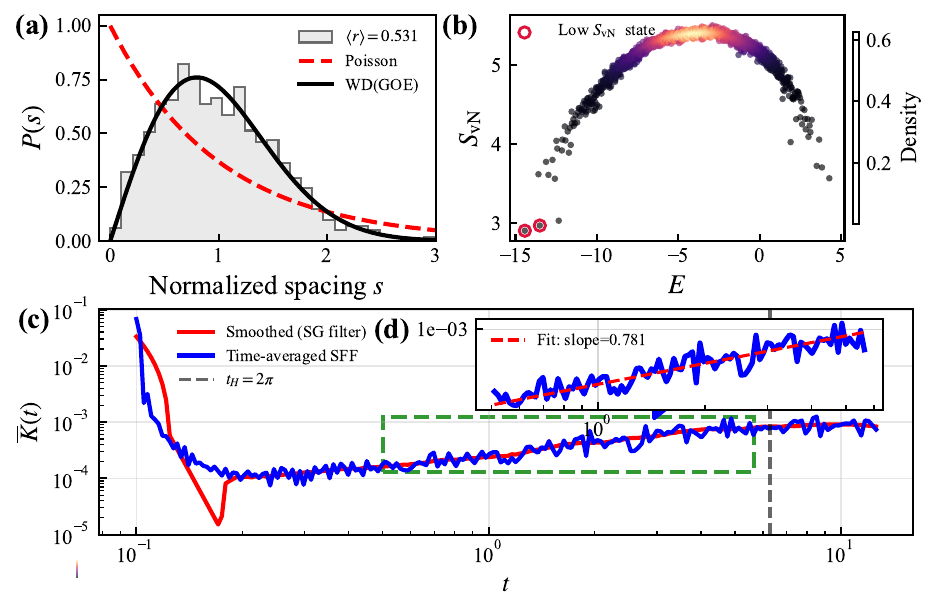}
        \caption{
Spectral and entanglement analysis for the $y$-translation symmetric sector with parameters $\alpha= (W_x = -1 , W_y = 0  , I_A = 4 ,I_B = 16  ,I_C =  32 ,I_D = 20 )$ and momentum $\mathbf{k} = (0,\pi/2)$. (a) Level-spacing histogram $P(s)$ with unfolded eigenvalues, showing good agreement with the Wigner-Dyson (GOE) distribution (black solid line) and comparison to the Poisson distribution (red dashed line); the mean ratio $\langle r\rangle = 0.531$ is indicated. (b) Bipartite von Neumann entanglement entropy $S
_{vN}$ versus energy $E$, with the density estimated via a Gaussian KDE (colorbar) and low-entanglement states highlighted by red circles below the main thermal dome. (c) Time-averaged spectral form factor (SFF, blue) and smoothed SFF using Savitzky–Golay filter (red). (d) Zoomed-in view of the ramp region of the SFF (highlighted in green in panel c), showing a linear fit (red dashed line) with slope $0.781$, indicating partial level repulsion in the bulk spectrum.
}        \label{fig:svn2}
    \end{figure}
\end{center}
\end{widetext}

FIG. \ref{fig:svn2} shows the spectral and entanglement properties of the $k_y = \pi/2$ momentum subsector. Panel (a) displays the level-spacing distribution $P(s)$ of unfolded eigenvalues, which follows the Wigner–Dyson form for the bulk spectrum (black solid line), with the Poisson distribution shown for comparison (red dashed line). The average adjacent-gap ratio $\langle r\rangle \approx 0.53$ indicates appreciable level repulsion. Panel (b) presents the bipartite von Neumann entanglement entropy $S_{vN}$
 versus energy $E$, with the density estimated via a Gaussian kernel; the majority of eigenstates form a smooth, dome-shaped distribution, indicative of thermal behavior, and no low-entanglement outliers are observed. Panel (c) shows the time-averaged spectral form factor (SFF, blue) and its smoothed counterpart using a Savitzky–Golay filter (red), capturing the initial decay and onset of the ramp. Panel (d) provides a zoomed-in view of the ramp region (highlighted in green in panel c) with a linear fit (red dashed line) yielding a slope of $0.781$, quantifying the correlation growth in the bulk spectrum. These panels demonstrate that the $k_y=\pi/2$
 subsector exhibits spectral correlations characteristic of quantum chaos, while the eigenstates remain fully thermal, with no evidence for quantum scar states.

Extending this analysis to all four momentum subsectors along the $y$-direction $(k_y=0,\pi/2,\pi,3\pi/2)$, each of equal dimension (2190) within the largest connected Krylov component (dimension 8760), we find that the graph structure induced by nonzero off-diagonal Hamiltonian elements is essentially identical: a single connected component, comparable mean degree ($\langle z\rangle \approx 8$), small relative degree fluctuations ($\mathrm{std}(z) / \langle z\rangle \approx 0.26$) and exactly one zero mode in the graph Laplacian. Despite this structural equivalence, the spectral statistics vary: $k_y = \pi/2$  and $k_y = 3\pi/2$ display Wigner–Dyson-like correlations ($\langle r\rangle \approx 0.53$), while 
$k_y =0$  and $k_y = \pi$ exhibit weaker correlations ($\langle z\rangle \approx 0.42-0.45$), yet all sectors show dome-shaped entanglement entropy and SFF behavior indicative of thermalization. Notably, no low-entanglement outliers are observed in any sector, suggesting the absence of quantum scar states. These results illustrate that momentum projection affects level statistics, yet the underlying Hilbert-space connectivity ensures that eigenstates remain thermal.

\textbf{\textit{Case}-III:} In the sector with both the $x$- and $y$-direction translational symmetry resolved, extracting the largest connected Krylov component (total dimension 23124) yields sixteen momentum subsectors with $k_x,k_y \in \{0,\pi/2,\pi,3\pi/2\}$, whose dimensions range from 1440 to 1458.

\begin{widetext}
\begin{center} 
    \begin{figure}[H]
        \centering
        \includegraphics[width=0.8\textwidth]{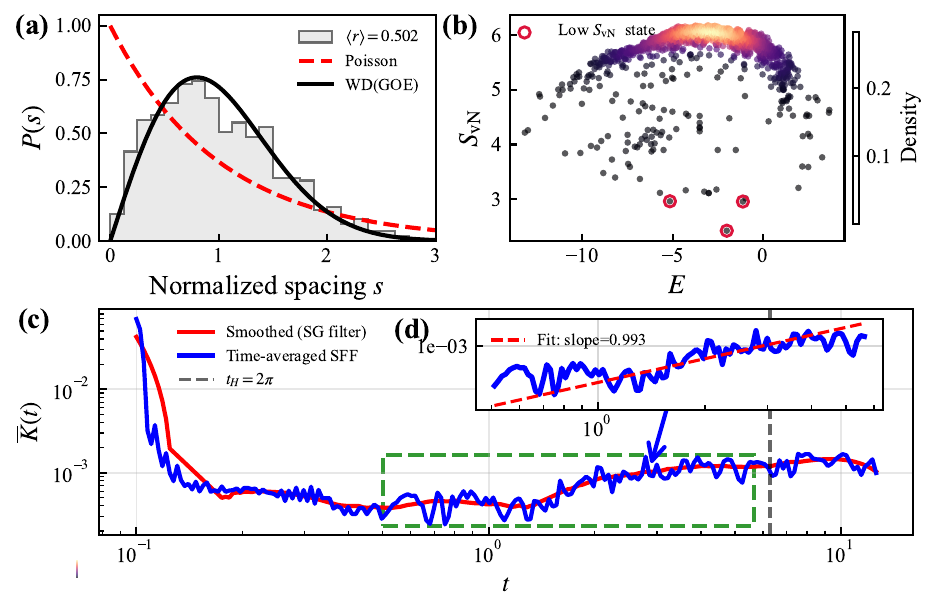}
        \caption{
Spectral and entanglement analysis for the $x$and$y$-translation symmetric sector with parameters $\alpha=(W_x = 0 , W_y = 0  , I_A = -16 ,I_B = -16  ,I_C =  0 ,I_D = -16 )$ and momentum $\mathbf{k} = (3\pi/2,3\pi/2)$. (a) Level-spacing histogram $P(s)$ with unfolded eigenvalues, showing good agreement with the Wigner-Dyson (GOE) distribution (black solid line) and comparison to the Poisson distribution (red dashed line); the mean ratio $\langle r\rangle = 0.502$ is indicated. (b) Bipartite von Neumann entanglement entropy $S
_{vN}$ versus energy $E$, with the density estimated via a Gaussian KDE (colorbar) and low-entanglement states highlighted by red circles below the main thermal dome. (c) Time-averaged spectral form factor (SFF, blue) and smoothed SFF using Savitzky–Golay filter (red). (d) Zoomed-in view of the ramp region of the SFF (highlighted in green in panel c), showing a linear fit (red dashed line) with slope $0.993$, indicating partial level repulsion in the bulk spectrum.
}        \label{fig:svn3}
    \end{figure}
\end{center}
\end{widetext}
FIG. \ref{fig:svn3} shows the spectral and entanglement properties of a representative momentum subsector, $(k_x ,k_y )=(3\pi/2,3\pi/2)$, within the $x$- and $y$-translation symmetric sector. Panel (a) displays the level-spacing histogram $P(s)$ of unfolded eigenvalues, which follows the Wigner–Dyson form for the bulk spectrum (black solid line), with the Poisson distribution shown for comparison (red dashed line); the average adjacent-gap ratio $\langle r\rangle \approx 0.502$ indicates significant level repulsion. Panel (b) presents the bipartite von Neumann entanglement entropy $S_{vN}$ versus energy $E$, with the density estimated via a Gaussian kernel; the majority of eigenstates form a smooth, dome-shaped distribution, and a few low-entanglement states appear below the main thermal dome, consistent with quantum scar states. Panels (c) and (d) show the time-averaged spectral form factor (SFF) and its zoomed-in ramp region, respectively, revealing thermalization in the bulk spectrum. These observations confirm that this subsector exhibits signatures of chaotic spectral correlations alongside fully thermal eigenstates and isolated scars.

Extending this analysis to all 16 momentum subsectors, we find that the graph structure induced by nonzero off-diagonal Hamiltonian elements is essentially identical across subsectors: each forms a single connected component, and the graph Laplacian spectrum contains exactly one zero mode, confirming that each Krylov component is connected. Within this framework, four subsectors—$(k_x ,k_y )=(\pi/2,\pi/2),(\pi/2,3\pi/2),(3\pi/2,\pi/2),(3\pi/2,3\pi/2)$—exhibit Wigner–Dyson-like level repulsion ($\langle r\rangle \approx 0.502$), whereas the remaining twelve subsectors display weaker correlations ($\langle r\rangle \approx 0.4-0.45$). Remarkably, in all subsectors, the entanglement entropy forms a dome-shaped distribution, with low-entanglement states appearing below the main dome in sectors hosting scar states. These results illustrate that momentum projection leads to varied spectral correlations, yet thermalization persists across all subsectors, with scar states appearing only in selected ones.

Across the three representative sectors analyzed, we find a consistent pattern: individual momentum subsectors exhibit variations in spectral statistics, with some showing Wigner–Dyson-like level repulsion ($\langle r\rangle \approx 0.5$) and others weaker correlations ($\langle r\rangle \approx 0.4-0.45$). Remarkably, in all subsectors and across all sectors, the many-body eigenstates form smooth, dome-shaped energy-resolved entanglement entropy distributions, with low-entanglement scar states appearing only in a subset of subsectors. Time-averaged spectral form factor analysis further confirms that the bulk of the spectrum behaves thermally, regardless of momentum-dependent variations in level statistics. These observations demonstrate that, while momentum projection can strongly modulate spectral correlations, the underlying Hilbert-space connectivity ensures that the majority of eigenstates remain thermal, and scar states appear only as isolated deviations. Taken together, the results from the three sectors provide a comprehensive picture of constrained quantum dynamics: high local constraints can generate sector-selective nonergodicity, yet most of the Hilbert space satisfies the eigenstate thermalization hypothesis, highlighting the coexistence of thermal behavior and scar states in highly constrained lattice models.

\textbf{\textit{Conclusions.—}}
In summary, our analysis of three representative sectors of the height-conserving quantum dimer model demonstrates sector-dependent spectral statistics, distinct from both Poisson and fully Wigner–Dyson statistics. By resolving the spectrum into momentum subsectors, we show that this behavior persists within the dominant connected Krylov component of each sector, and arises from the interplay of local dimer constraints, the conserved height field, and lattice symmetries. While spectral correlations vary across momentum subsectors—some displaying Wigner–Dyson-like level repulsion and others exhibiting weaker correlations—the majority of many-body eigenstates in all sectors form smooth, dome-shaped, energy-resolved entanglement bands, indicative of thermalization, with low-entanglement scar states appearing only in selected subsectors. Time-averaged spectral form factor analysis further confirms the thermal nature of the bulk spectrum. Thus, strong local constraints alone can lead to momentum-dependent spectral statistics while thermalization of eigenstates remains robust—a clear illustration that spectral correlations and ETH need not follow the same universal behavior. This establishes a concrete example of constrained quantum chaos, where the standard diagnostics can appear to disagree, yet the system still supports thermalization in its eigenstates, consistent with recent studies in related quantum dimer models \cite{Lan2017,Wildeboer2021}.

\textbf{\textit{Acknowledgments.}}
The authors thank Dr. Zhicheng Yang for helpful discussions. This work is  supported by  the Scientific Research Project of
Universities in Anhui Province (Grant No. 2024AH040216), the Scientific Research Project (No.WU2025B011) and the Start-up Funding of Westlake University.

\textit{Data availability}. All data are available from the authors upon reasonable request.

\bibliography{ref}
\end{document}